\documentclass[aps,prl,twocolumn,superscriptaddress]{revtex4-1}

\usepackage[latin9]{inputenc}
\usepackage[T1]{fontenc} 
\usepackage{lmodern} 
\usepackage{color}
\usepackage[colorlinks,urlcolor=blue,citecolor=blue,linkcolor=blue]{hyperref}
\usepackage{verbatim}
\usepackage{float}
\usepackage{amsmath}
\usepackage{amssymb}
\usepackage{graphicx}
\usepackage{epstopdf}
\usepackage{esint}
\usepackage{CJK}
\usepackage{braket}
\begin{document}
\begin{CJK*}{UTF8}{gbsn} 
\title{Probing the Interface of a Phase-Separated State in a Repulsive Bose-Fermi Mixture}


\author{Rianne~S.~Lous}
\author{Isabella~Fritsche}
\author{Michael~Jag}
\altaffiliation{Present address: LENS and Dipartimento di Fisica e Astronomia, Universit\`a di Firenze, 50019 Sesto Fiorentino, Italy}
\author{Fabian~Lehmann}
\affiliation{Institut f\"ur Quantenoptik und Quanteninformation (IQOQI), \"Osterreichische Akademie der Wissenschaften, 6020 Innsbruck, Austria}
\affiliation{Institut f\"ur Experimentalphysik, Universit\"at Innsbruck, 6020 Innsbruck, Austria}
\author{Emil~Kirilov}
\affiliation{Institut f\"ur Experimentalphysik, Universit\"at Innsbruck, 6020 Innsbruck, Austria}
\author{Bo~Huang (黄博)}
\email{Bo.Huang@uibk.ac.at}
\affiliation{Institut f\"ur Quantenoptik und Quanteninformation (IQOQI), \"Osterreichische Akademie der Wissenschaften, 6020 Innsbruck, Austria}
\author{Rudolf~Grimm}
\affiliation{Institut f\"ur Quantenoptik und Quanteninformation (IQOQI), \"Osterreichische Akademie der Wissenschaften, 6020 Innsbruck, Austria}
\affiliation{Institut f\"ur Experimentalphysik, Universit\"at Innsbruck, 6020 Innsbruck, Austria}


\date{\today}
\pacs{to be checked 34.50.Cx, 67.85.Lm, 67.85.Pq, 67.85.Hj}
\begin{abstract}
We probe the interface between a phase-separated Bose-Fermi mixture consisting of a small Bose-Einstein condensate of $^{41}$K residing in a large Fermi sea of $^6$Li. We quantify the residual spatial overlap between the two components by measuring three-body recombination losses for variable strength of the interspecies repulsion. A comparison with a numerical mean-field model highlights the importance of the kinetic energy term for the condensed bosons in maintaining the thin interface far into the phase-separated regime. Our results demonstrate a corresponding smoothing of the phase transition in a system of finite size.

\end{abstract}

\pacs{}

\maketitle
\end{CJK*}


Multicomponent systems and materials are ubiquitous in nature and technology. The interactions between the different constituents and the ways in which they coexist are essential for understanding the general properties of such systems. Repulsive interactions between different components can induce phase transitions to spatially separated states. The effects of phase separation appear in a wide range of different systems such as alloys, combinations of different liquids, colloids, polymers, glasses and biological systems. In a phase-separated state, the interaction between the components no longer takes place in the bulk but is restricted to the thin interface where the constituents still maintain some residual overlap. The physics of this interface has therefore attracted a great deal of attention in many different fields, e.g.\ in liquid-liquid systems~\cite{davis1996smo, hansen2013tos}. However, since the interaction takes place in a very small volume, it is generally much more difficult to obtain experimental information from these systems as compared to systems in which the components are mixed.

Quantum fluids exhibit a great wealth of phenomena related to phase separation. Early experiments with cryogenically cooled liquid helium have shown phase separation in mixtures of the bosonic isotope $^{4}$He and the fermionic $^{3}$He~\cite{Ebner1971tlt}. This effect has found an important technological application in the working principle of dilution refrigerators~\cite{Das1965aro, Pobell2007mam}. Ultracold gases, in particular, mixed-species systems have opened up many intriguing experimental possibilities to study phases of multicomponent quantum matter~\cite{Bloch2008mbp}. The large experimental toolbox includes a variety of available bosonic and fermionic constituents, a superb level of control of confinement, and a wide tunability of interactions~\cite{Chin2010fri}. Phase separation has been studied extensively in degenerate Bose-Bose mixtures~\cite{Papp2008tmi, Tojo2010cps, Mccarron2011dsb, Stamperkurn2013sbg, Wacker2015tds, Wang2016ads, Lee2016psa}, where interactions are dominated by mean-field potential energies. The situation becomes more complicated when fermionic constituents are involved, as strong repulsion on the scale of the Fermi energy is required to observe phase separation. Superfluid fermionic mixtures~\cite{Shin2008pdo} and repulsive atomic Fermi gases~\cite{Valtolina2017etf} are examples of intriguing phase-separation effects. In a broad sense, mixtures involving fermionic constituents are promising candidates for realizing new phases, e.g.,\ in Fermi-Fermi systems~\cite{Liu2003igs, Iskin2006tsf, Parish2007pfc, Baranov2008spb, Baarsma2010pam, Wang2017eeo} and in Bose-Fermi systems~\cite{Powell2005dot, Suzuki2008psa, Marchetti2008psa, Ludwig2011qpt, Bertaina2013qmc, Kinnunen2015iii}. 

In this Letter, we consider a Bose-Fermi model system that undergoes phase separation and study the interface between the constituents. We produce a Bose-Einstein condensate (BEC) of $^{41}$K atoms in a large Fermi sea of $^{6}$Li, and we use an interspecies Feshbach resonance for controlling the repulsive interaction. We characterize the overlap between the species by measuring three-body recombination losses and thus probe the thin interface between both components. By comparing the experimental results with theoretical model calculations, we demonstrate the importance of the kinetic energy of the condensed bosons at the thin interface.

 \begin{figure}
\includegraphics[clip,width=\columnwidth]{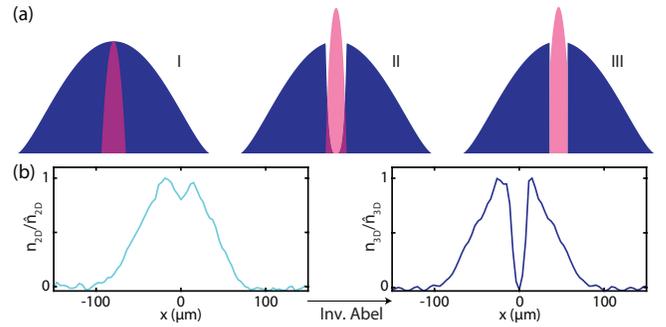}%
 \caption{Emergence of phase separation. (a) Schematic density profiles for bosons (magenta) and fermions (blue) for an increasing repulsive interaction. The densities are normalized to the corresponding peak value without an interaction. Note that in reality the boson peak density is a factor of 40 larger than the fermion peak density. (b) Experimentally observed normalized column density of a cut through the fermionic cloud and normalized reconstruction of the corresponding radial density profile using the inverse Abel transformation.}
\label{fig:idea}
 \end{figure}
Figure~\ref{fig:idea}(a) illustrates the onset of phase separation with an increasing interspecies repulsion, showing the density profiles of a small-sized BEC coexisting with a large Fermi sea in a harmonic trap. The main conditions and criteria for phase separation in such Bose-Fermi mixtures have been theoretically introduced in Refs.~\cite{Molmer1998bca, Viverit2000ztp, Roth2002sas}. For a vanishing interspecies interaction, the independent spatial profiles of the clouds show maximum overlap [I in Fig.~\ref{fig:idea}(a)]. With an increasing repulsion, the density of the lithium atoms in the center of the trap decreases, the BEC is compressed, and the spatial overlap between the clouds is reduced (II).  For strong repulsive interactions, the two clouds undergo phase separation (III), and the bosons reside at the center of the trap, forming a hole in the Fermi sea.

We can observe the depletion in the center of the Fermi sea by imaging the $^{6}$Li cloud. As Fig.~\ref{fig:idea}(b) shows, we observe a small dip in the radial column density profile taken from a thin slice of the fermion cloud. These data were taken under similar conditions as our main data presented later~\footnote{The thin slice is taken from a typical absorption image of the Li cloud with a time of flight of 2\,ms and at $a_\mathrm{bf}\approx 1480a_0$}. The hole in the fermion density becomes more visible when reconstructing the fermionic radial density profile using the inverse Abel transformation [Fig.~\ref{fig:idea}(b)]. We see an essentially complete depletion of the fermionic density in the center, which indicates a significant reduction of the overlap with the BEC. A quantitative analysis of the physics at the interface is obstructed by the limited signal-to-noise ratio of the image, the small size of the overlap region compared to our imaging resolution, and the high optical density of the trapped cloud. Note that strong indications of phase separation in a Bose-Fermi mixture have been observed in earlier experiments on mixtures of 
$^{87}$Rb and $^{40}$K~\cite{Ospelkaus2006toh, Zaccanti2006cot}, but these experiments did not provide quantitative information on the overlap reduction.

Here, we introduce an alternative approach to study the spatial overlap between the two species. Our observable is the boson-boson-fermion three-body recombination loss from the trap.
We assume that all losses can be attributed to three-body processes, since two-body losses are energetically suppressed when both atomic species are in their lowest internal substates. In our system, decay processes of three $^{41}$K atoms (three identical bosons) occur at a very low rate, since the intraspecies scattering length $a_\mathrm{bb}=60.9a_0$~\cite{Tiemannpriv}, with $a_0$ being the Bohr radius, is small compared with the interspecies scattering length $a_\mathrm{bf}$ in the range of interest. On the other hand, recombination processes involving one $^{41}$K atom and two $^6$Li atoms (one boson and two identical fermions) are Pauli suppressed~\cite{Esry2001tlf}. At a large interspecies scattering length, this leaves the recombination events of two $^{41}$K atoms with one $^6$Li atom as the dominant three-body decay mechanism. 

A favorable property of our system is the fact that the BEC is much smaller than the fermion cloud and occupies a very small volume within the Fermi sea. Thus, the BEC can cause only a local perturbation of the Fermi sea with a negligible effect on the global scale. This scenario enables a description in terms of a \textit{fermionic reservoir approximation} (FRA), which assumes a homogeneous environment characterized by a constant Fermi energy $E_\mathrm{F}$ and considerably simplifies our study of the overlap.

In the zero-temperature limit, where a pure BEC is formed, the bosonic atom loss can be related to the overlap integral as
\begin{equation}
\dot{N}= -\frac{1}{2}\,L_\mathrm{3}\int  n_\mathrm{f}\,n_\mathrm{b}^2\, \mathrm{d}V, 
\label{dNfull}
\end{equation}
where $N$ is the total number of bosons and $n_\mathrm{b}$ and $n_\mathrm{f}$ represent the position-dependent number densities of the bosons and fermions, respectively. The parameter $L_\mathrm{3}$ is the three-body loss coefficient, and the symmetry factor $1/2$ results from the suppression of thermal bunching in a BEC~\cite{Kagan1985eob, Burt1997cca, Soding1999tbd, Haller2011tbc} for a process involving two identical bosons. The $L_\mathrm{3}$ coefficient can be determined as a function of $a_\mathrm{bf}$ in a standard way~\cite{Weber2003tbr, Ulmanis2016utb} using a noncondensed cloud instead of a BEC. In this case, the interspecies repulsion can be neglected, and the density profiles of the bosons and the fermions are well known.



In order to characterize the effect of the boson-fermion interaction on the spatial overlap between the BEC and the Fermi sea, we define the overlap factor 
 \begin{equation}
\Omega\equiv\frac{\int{ n_\mathrm{f}\,n_\mathrm{b}^2\, \mathrm{d}V}}{\int\tilde{n}_\mathrm{f}\,\tilde{n}_\mathrm{b}^2\, \mathrm{d}V}
\label{Omega} 
\end{equation}
as the three-body density integral normalized to the case of vanishing interspecies interaction ($a_\mathrm{bf}=0$), where $\tilde{n}_\mathrm{f}\,(\tilde{n}_\mathrm{b})$ is the fermionic (bosonic) noninteracting density.

The overlap integral for the case of a vanishing interspecies interaction, $\int\tilde{n}_\mathrm{f}\,\tilde{n}_\mathrm{b}^2\, \mathrm{d}V$, can be calculated analytically based on two approximations. First, we apply the FRA and replace $\tilde{n}_\mathrm{f}$ by its peak value $\hat{n}_\mathrm{f}$, which as a constant factor can be taken out of the integral. Second, for a not too small BEC, we can apply the Thomas-Fermi approximation and solve $\int \tilde{n}_\mathrm{b}^2\, \mathrm{d}V$ as $\frac{4}{7}\, N_\mathrm{b}\, \hat{n}_\mathrm{b}$, with $\hat{n}_\mathrm{b}$ the peak density of the BEC. Finally, with the overlap integral for the interacting case given by Eq.~(\ref{dNfull}), the overlap factor can be experimentally obtained as
\begin{equation}
\Omega=\frac{7}{2\,\hat{n}_\mathrm{f}\,\hat{n}_\mathrm{b}}\, \frac{\gamma}{L_\mathrm{3}},
\label{Omega2}
\end{equation}
where we introduce the normalized loss rate ${\gamma = - \dot{N}/N}$ as the experimental observable extracted from measuring the atom loss in a BEC. 

For our experiments, we prepare an ultracold Bose-Fermi mixture of typically $10^4$ K and $10^5$ Li atoms in a cigar-shaped, crossed-beam optical dipole trap with a wavelength of 1064\,nm and an aspect ratio of 1:7. The preparation procedures are similar to those described in Ref.~\cite{Lous2017toa} and earlier work on $^{6}$Li-$^{40}$K mixtures~\cite{Spiegelhalder2009cso,Trenkwalder2011heo,Kohstall2012mac, Jag2014ooa, Cetina2016umb}. In addition, we employ a laser cooling scheme for lithium using the $D$1 line~\cite{Grier2013les,Burchianti2014eao,Fritsche2015Msc}, which provides improved starting conditions, and we take advantage of an alternative evaporative cooling approach~\cite{Burchianti2014eao, supp}. 

A key ingredient of our experiment is the Feshbach resonance (FR) near 335\,G~\cite{supp, Hannapriv, Tiemannpriv, Wu2011sii, Lous2017PhD}, between the lowest spin states of the two species. The scattering length can be varied by magnetic field tuning according to $a_\mathrm{bf}=a_\mathrm{bg}\left[1-\Delta/(B-B_0)\right]$~\cite{Chin2010fri}, where $\Delta=0.949\,$G, $a_\mathrm{bg}=60.9a_0$ and $B_0=335.057(1)\,$G. The FR center $B_0$ somewhat depends on the optical trap intensity because of a light shift effect~\cite{supp} and can be experimentally determined by radio-frequency spectroscopy. The other parameters are obtained from scattering models~\cite{Hannapriv, Tiemannpriv, supp}. 

To obtain the critical interspecies scattering length for the onset of phase separation, we employ the FRA together with the results of Ref.~\cite{Viverit2000ztp}. This yields the condition 
\begin{equation}
a_\mathrm{bf}>1.15\sqrt{a_\mathrm{bb}/k_\mathrm{F}},
\label{eq:ps}
\end{equation}
where $k_\mathrm{F}= (6 \pi^2 \hat{n}_\mathrm{f})^{1/3}$ is the Fermi wave number, corresponding to $E_\mathrm{F} = \hbar^2 k_\mathrm{F}^2 / (2 m_\mathrm{f})$ with $m_\mathrm{f}$ the mass of the fermions. For our typical experimental conditions ($\hat{n}_\mathrm{f}\approx1.2\times10^{12}\,\mathrm{cm}^{-3}$), it gives a moderate value for the critical scattering length of about $600a_0$. This is well within our tuning range and allows us to explore the entire scenario from weak to strong repulsion, reaching far into the phase-separated regime. 

 \begin{figure}
\includegraphics[clip,width=\columnwidth]{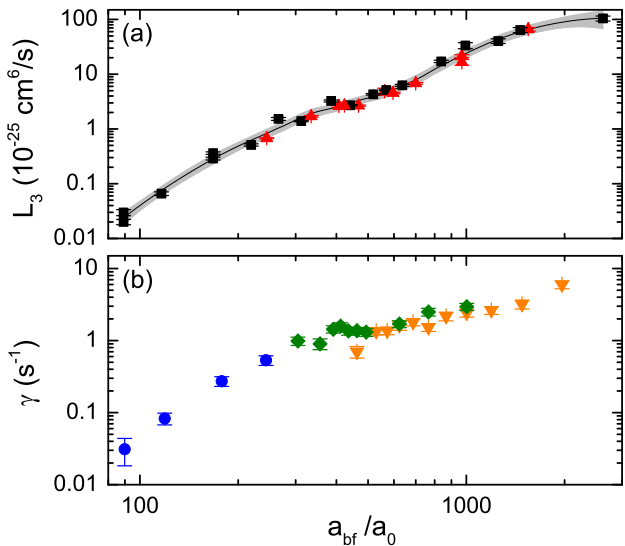}%
 \caption{Loss measurements on noncondensed and condensed bosonic $^{41}$K clouds in a $^6$Li Fermi sea. The error bars represent 1$\sigma$ fit uncertainties. (a) Three-body loss coefficient $L_3$ for $T=440$ (set $A$1: squares) and 240\,nK (set $A$2: triangles). The solid curve is an interpolation from applying a smoothing method~\cite{supp}, with the gray-shaded area representing the corresponding 95\% confidence band. (b) Normalized loss rate $\gamma$ of the total atom number of a partially condensed bosonic cloud for data sets $B$1-$B$3 (inverse triangles, diamonds, and circles, respectively).}
\label{fig:L3loss}
 \end{figure}
We first present our measurements of $L_3$, which were obtained with noncondensed samples of $^{41}$K in a degenerate Fermi sea of $^{6}$Li at about $0.2T_\mathrm{F}$, with $T_\mathrm{F}$ the Fermi temperature. From the measured decay curves we obtain the $L_3$ values that are shown in Fig.~\ref{fig:L3loss}(a). The K samples are prepared close to degeneracy at two different temperatures with a typical fermion peak density of $\hat{n}_\mathrm{f}\approx4.5\times10^{12}\,$cm$^{-3}$. In one set of measurements (set $A$1)~\cite{supp}, we have $T=440\,$nK, corresponding to $T/T_\mathrm{c} = 1.7$ with $T_\mathrm{c}$ the critical temperature for condensation. In the other set ($A$2), we have $T=240\,$nK, corresponding to $T/T_\mathrm{c} \approx 1$. By applying a smoothing method~\cite{supp}, we interpolate between the data points and  obtain $L_3$ for any $a_\mathrm{bf}$ between 80 and 2100$a_0$. Our results on $L_3$ show the expected strong increase with $a_\mathrm{bf}$, while the broad dent around $600a_0$ may point to an Efimov-related feature~\cite{, Kraemer2006efe, Johansen2017tuo}.

Second, we present the boson loss rate $\gamma$ in a degenerate Bose-Fermi mixture at various interaction strengths. Typically, we have $2.9\times10^4$ K atoms with a 50\% condensate fraction in a Fermi sea of $1.4\times10^5$ Li atoms with a peak density of $\hat{n}_\mathrm{f}= 1.2\times10^{12}\,$cm$^{-3}$ and a temperature of $\sim 0.13T_\mathrm{F}$. The sample is first prepared at 200\,mG below $B_0$, and then the magnetic field is changed in a near-adiabatic ramp of 2\,ms to the specific field on the repulsive side of the FR, where we observe the loss of the K atoms for various hold times. We fit the initial decay of the total atom number with a linear curve and determine the normalized loss rate $\gamma$~\cite{supp}. 
 Figure~\ref{fig:L3loss}(b) shows the corresponding data points, which were recorded in three sets ($B$1-$B$3)~\cite{supp} with slightly varying parameters. 

 
With the normalized loss rate $\gamma$ and the three-body recombination coefficient $L_3$, we can now quantify the spatial overlap. In a real experiment, two complications arise that require an extension of our model beyond Eq.~\eqref{Omega2}. First, at a finite temperature, we have only a partial BEC and the presence of the thermal component plays a significant role in the observed loss. Second, there is the possibility of observing secondary loss, where a short-lived LiK dimer, produced in a first recombination, recollides with another K atom, and therefore this leads to additional loss~\cite{supp}. This process is likely to happen for the dense BEC but negligible for the thermal K cloud. To take both effects into account, we extend Eq.~(\ref{dNfull}) and include all loss contributions:
\begin{equation}
\dot{N}= - L_3\int{n}_\mathrm{f}\, \left(\frac{1}{2} \alpha\,n_\mathrm{b}^2+\alpha\, n_\mathrm{b}\,n_\mathrm{t}+ n_\mathrm{t}^2\right)\, \mathrm{d}V,
\end{equation}
 where ${n}_\mathrm{t}$ represents the thermal bosonic density and $\alpha$ is a factor that takes into account secondary loss. In our case, we assume $\alpha=3/2$~\cite{supp}. The density integral consists of three terms, which describe the loss caused by one fermion and two bosons. The bosons can either be two atoms from the BEC, one from the BEC and one from the noncondensed component, or two from the noncondensed bosonic cloud. Within the FRA and the Thomas-Fermi approximation, these integrals can be calculated, and an effective overlap factor results from an extension of Eq.~(\ref{Omega2}) as 
\begin{align}
\label{Omegaeff}
\Omega_\mathrm{eff}=\frac{1}{\hat{n}_\mathrm{f}\, \left[\frac{2}{7}\,\alpha\,\hat{n}_\mathrm{b}\,\beta   +\alpha\, \hat{n}_\mathrm{t} \beta  + \frac{1}{\sqrt{8}}\,\hat{n}_\mathrm{t}(1-\beta)\right]}\,\frac{\gamma}{L_3}, 
\end{align}
where $\beta$ is the BEC fraction and $\hat{n}_\mathrm{t}$ the peak density of a thermal Bose gas, as given by ${\hat{n}_\mathrm{t}=}{\left[m_\mathrm{b}\,\bar{\omega}_\mathrm{b}^2/(2\pi k_\mathrm{B}T)\right]^{3/2}{(1-\beta)\,N}}$, with $\bar{\omega}_\mathrm{b}$ being the geometrically averaged trap frequency of the bosons, $m_\mathrm{b}$ their mass, and $T=T_\mathrm{c}(1-\beta)^{1/3}$~\cite{Lous2017toa,supp}.

\begin{figure}
\includegraphics[clip,width=\columnwidth]{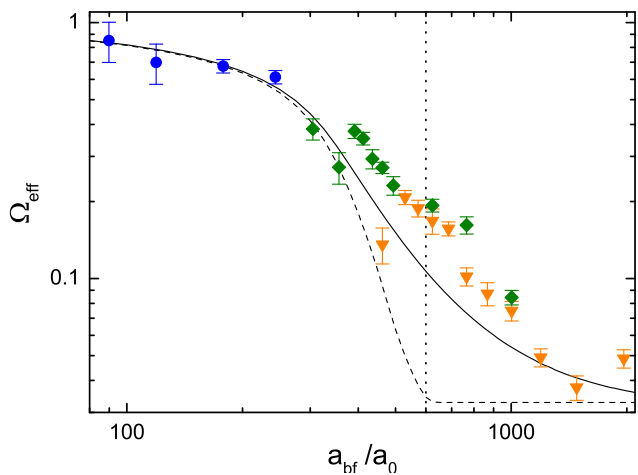}%
 \caption{Effective overlap factor versus Bose-Fermi scattering length for data sets $B$1-$B$3 (inverse triangles, diamonds, and circles, respectively). The error bars reflect the statistical uncertainties of $\gamma$. The vertical dotted line shows the phase-separation point as predicted by Eq.~(\ref{eq:ps}). The solid line shows the results of our full numerical calculation (see the text) and the dashed line our results obtained within the Thomas-Fermi approximation.}
\label{fig:Omega} 
 \end{figure}

Figure~\ref{fig:Omega} shows the values of $\Omega_\mathrm{eff}$ that result from the data in Fig.~\ref{fig:L3loss}.  We qualitatively distinguish three regions. Below $a_\mathrm{bf}\approx 250a_0$, the values are close to one, and there seems to be a downward trend for $\Omega_\mathrm{eff}$ with increasing $a_\mathrm{bf}$. Then, as $a_\mathrm{bf}$ further increases to about $1000a_0$, the spatial overlap drastically decreases to a small value of about $0.04$. For larger scattering lengths, $\Omega_\mathrm{eff}$ tends to remain at this small value. According to Eq.~(\ref{eq:ps}), phase separation is expected to happen at ${\sim600a_0}$ (vertical dotted line). In contrast, we observe that beyond this point a considerable spatial overlap remains, which then smoothly decreases with a further increasing scattering length. The observed behavior does not reveal any discontinuity related to a phase transition.

To interpret the observed behavior of $\Omega_\mathrm{eff}$, we construct a numerical mean-field model~\cite{supp, Huang2017} which allows us to calculate the density distributions for an interacting Bose-Fermi mixture at a zero temperature for our typical experimental parameters~\footnote{In our model, we consider an atom-atom mixture, neglecting any molecular component as the LiK Feshbach molecules are short-lived and decay rapidly.}. Our model starts from the energy functional of the mixture as given by Refs.~\cite{Imambekov2006bot,Trappe2016gsd}, and we use imaginary time evolution to vary the BEC and the fermionic densities and to minimize the energy functional. At the end, the evolution gives the static solution of $n_\mathrm{f}$ and $n_\mathrm{b}$ at a zero temperature. Since we have a partial BEC, we additionally take into account the thermal bosonic density $n_\mathrm{t}$ including bosonic enhancement effects~\cite{supp}. 
With these density distributions, we numerically calculate the overlap integrals and the effective overlap factor $\Omega_\mathrm{eff}$.

The results of our numerical model are represented in Fig.~\ref{fig:Omega} by the dashed and solid curves. For the dashed curve, the densities are obtained within the Thomas-Fermi approximation. The results indeed show a rapid decrease of $\Omega_\mathrm{eff}$ until the onset of phase separation at about $600a_0$, as given by Eq.~(\ref{eq:ps}). Then, in a fully phase-separated regime, a plateau is reached where only the thermal bosonic component can lead to loss. Evidently, this theoretical behavior is not consistent with the experimental data points.  A notably smoother decrease of $\Omega_\mathrm{eff}$ results from our numerical model (solid line in Fig.~\ref{fig:Omega}), when we consider the full energy functional which includes the kinetic energy of the BEC as well as the much weaker density gradient correction from the Fermi gas~\cite{Imambekov2006bot}. Within the residual uncertainties of our method~\cite{supp}, this model reproduces the observed behavior very well.

Our results show that the kinetic energy term prevents the BEC density from changing abruptly. This plays an essential role in smoothing the density profiles of the separated components near the interface and, thus, in maintaining the residual spatial overlap. Accordingly, the relevant length scale that determines the thickness of the interface layer corresponds to the BEC healing length~\cite{Dalfovo1999tob}, which for our present conditions can be estimated to $\xi = \left(8 \pi\, \hat{n}_\mathrm{b}\,a_\mathrm{bb}\right)^{-1/2} \approx 0.50\,\mu$m. This length scale can be compared with the shortest macroscopic length scale of the system, which in our case is the radial size of the BEC of a few micrometers. 
The measured overlap factor can be understood as the volume ratio of the interface layer and the whole BEC, and the smoothing of the phase transition can thus be interpreted as a consequence of the finite size of the system~\cite{Binder1984fss, Brezin1985fse}.

The basic idea of our method to probe the interface between spatially separated components may be generalized to many other situations of interest. The working principle just relies on a mechanism that selectively addresses the region where the different components mix. While in our case three-body recombination served this purpose, one may also apply photoassociative or radio-frequency-induced processes to stimulate loss or state-transfer processes. 

The interface between two quantum fluids is a topic of broad interest yet largely unexplored in quantum gases. We speculate that future studies could focus on the role of quantum fluctuations, the two-dimensional character of the thin interface layer, and testing the validity of the mean-field approach. Unwinding the microscopic nature underlying the interface may give access to new phenomena such as Andreev bound states~\cite{Loefwander2001abs, Masatoshi2017tsa}, familiar in superconductor physics. Concerning the phase-separated Bose-Fermi mixture, it would be natural to go beyond the static properties and to investigate the dynamics of the mixture. We expect a strong impact of phase separation on collective oscillation modes~\cite{VanSchaeybroeck2009tps, Maruyama2013lat} and on the behavior of the system after a quench~\cite{Will2015ooc}. 

\begin{acknowledgments}
We acknowledge valuable discussions with M.~Baranov and D.~Yang on the theoretical model, with E.~Tiemann and P.~S.~Julienne on the scattering properties, and with A.~Turlapov on general topics. We thank J.~T.~M.~Walraven for fruitful insights and our new team members C.~Baroni, A.~Bergschneider, \mbox{T.~W.~Grogan} and T.~\"Ottl for comments on the manuscript. We acknowledge support by the Austrian Science Fund FWF within the Spezialforschungsbereich FoQuS (F4004-N23) and within the Doktoratskolleg ALM (W1259-N27).
\end{acknowledgments}


\bibliographystyle{apsrev4-1}
\bibliography{ultracold,PhaseSeparationNewrefs}

\end{document}